# X-ray induced grain boundary formation and grain rotation in $Bi_2Se_3$

Kento Katagiri[1,2,3*], Bernard Kozioziemski[4], Eric Folsom[4], Sebastian Göde[5], Yifan Wang[1,2,3], Karen Appel[5], Darshan Chalise[1,2,3], Philip K. Cook[6], Jon Eggert[4], Marylesa Howard[7], Sungwon Kim[8], Zuzana Konôpková[5], Mikako Makita[5], Motoaki Nakatsutsumi[5], Martin M. Nielsen[9], Alexander Pelka[10], Henning F. Poulsen[9], Thomas R. Preston[5], Tharun Reddy[1,2,3], Jan-Patrick Schwinkendorf[5], Frank Seiboth[11], Hugh Simons[9], Bihan Wang[12], Wenge Yang[12], Ulf Zastrau[5], Hyunjung Kim[8], and Leora E. Dresselhaus-Marais[1,2,3,4*]

[1]*Department of Materials Science and Engineering, Stanford University, Stanford, CA 94305, USA*

[2]*SLAC National Accelerator Laboratory, Stanford, CA 94025, USA*

[3]*PULSE Institute, Stanford University, Stanford, CA 94305, USA*

[4]*Lawrence Livermore National Laboratory, Livermore, CA 94550, USA*

[5]*European XFEL GmbH, Schenefeld 22869, Germany*

[6]*Institute for Physics and Materials Science, University of Natural Resources and Life Sciences (BOKU), Vienna 1190, Austria*

[7]*Nevada National Security Site, North Las Vegas, NV 89030, USA*

[8]*Department of Physics, Sogang University, Seoul 04107, Korea*

[9] *Department of Physics, Technical University of Denmark, Kgs. Lyngby 2800, Denmark*

[10]*Helmholtz-Zentrum Dresden-Rossendorf (HIBEF), Dresden D-01328, Germany*

[11]*Center for X-ray and Nano Science CXNS, Deutsches Elektronen-Synchrotron DESY, Hamburg 22607, Germany*

[12]*Center for High-Pressure Science and Technology Advanced Research, Shanghai 201203, China*

*Corresponding authors.
Email addresses:
kentok@stanford.edu (K. Katagiri),
leoradm@stanford.edu (L. Dresselhaus-Marais)

**Optimizing grain boundary characteristics in polycrystalline materials can improve their properties. Many processing methods have been developed for grain boundary manipulation, including the use of intense radiation in certain applications. In this work, we used X-ray free electron laser pulses to irradiate single-crystalline bismuth selenide ($Bi_2Se_3$) and observed grain boundary formation and subsequent grain rotation in response to the X-ray radiation. Our observations with simultaneous transmission X-ray microscopy and X-ray diffraction demonstrate how intense X-ray radiation can rapidly change size and texture of grains.**

Grain boundary engineering (GBE) is a technique to control and optimize grain boundaries (GBs) in materials to improve their properties such as mechanical strength, electrical conductivity, and corrosion resistance [1,2]. While many GBE processing methods are based on mechanical, thermal, chemical, or a combination of these processes [1,3,4], previous studies have shown that intense ion or X-ray radiation can also change GB characteristics [5-11]. Intense ion or X-ray beams on samples generate additional defects localized at GBs that can exhibit non-thermal migration effects and enhanced mobility as compared to classical GBs [11]. In nuclear reactors, the competition between radiation, extreme heating, and dose rates cause additional GB segregation that can degrade or strengthen structural metals [12]. Given the conflicting result on properties, understanding GB effects in radiation environments are critical to ensure their effective function in their intended application [13,14].

While displacive radiation damage by ions interacts with materials differently from non-displacive radiation by X-rays and gamma rays, their resulting material interactions have key similarities based on the relevant timescales and temperatures [15,16]. Uniquely, non-displacive radiation by X-rays offers the ability to directly characterize the progression of subtle GB changes during irradiation, due to recent advances in their generation [17,18]. X-ray free electron lasers (XFELs) offer femtosecond X-ray pulses with high peak brightness that can be used to both *pump* (*i.e.*, excite heating/damage in the sample) and *probe* (*i.e.*, observe) with high time resolutions. Also, the nano-focusing techniques recently developed for XFELs [19,20] and synchrotrons [21] can be used for local GBE, with simultaneous characterizations with nanometer precisions. Thus, the use of novel X-ray techniques to study how intense radiation changes grain and GB structures is an important new direction that has not been explored to date.

While GBE has been mainly studied for metals and alloys, its importance in other materials such as ceramics and semiconductors has been also recognized because of their industrial importance [1,22,23]. Especially, GBE to tailor the microstructure of thermoelectric materials, such as $Bi_2Se_3$ [24], can be useful for their applications as their thermoelectric properties are strongly determined by GBs [24,25] and grain orientation [26]. In this study, we used the high-intensity X-ray pulses generated at European XFEL to irradiate a bismuth selenide ($Bi_2Se_3$) single crystal and time-resolved the X-ray induced GB formation and subsequent grain rotations. Using simultaneous transmission X-ray microscopy (TXM) [27] and X-ray diffraction (XRD), we demonstrate how the heat associated with extreme X-ray absorption gives rise to GB formation and subsequent grain rotations. Although the spatial resolution of such X-ray measurements is generally inferior to transmission electron microscope (TEM) measurements [28-35], X-ray experiments at synchrotron and XFEL sources enable high temporal resolution measurements that are characteristic of the timescales relevant to radiation effects in materials [36-38].

This experiment was performed at the High Energy Density (HED) instrument of the European XFEL [40,41]. As shown in Fig. 1(a), we used the XFEL beam to simultaneously measure TXM and XRD. Though both TXM and XRD measurements can be done non-destructively, the X-ray intensity used in this work was tuned to be high enough to cause damage to the sample. We thus used the X-ray pulses both as the *pump* and the *probe* of the material pumped by the previous pulses. Using the femtosecond-duration X-ray pulses generated by the XFEL, its *in-situ* XRD can time-resolve the dynamics of structural changes such as lattice deformation, structural transformations, or changes to the orientations of different crystallites. *In-situ* TXM records the real-space intensity map of the X-ray beam transmitted through the sample. For TXM, the X-ray beam transmitted the sample is magnified by using X-ray lenses such as compound refractive lenses (CRLs) [43], to form a real-space image in the far field (*i.e.*, the X-ray image is collected at a position that is a long distance from the sample being measured) with sub-micrometer spatial resolution [44].

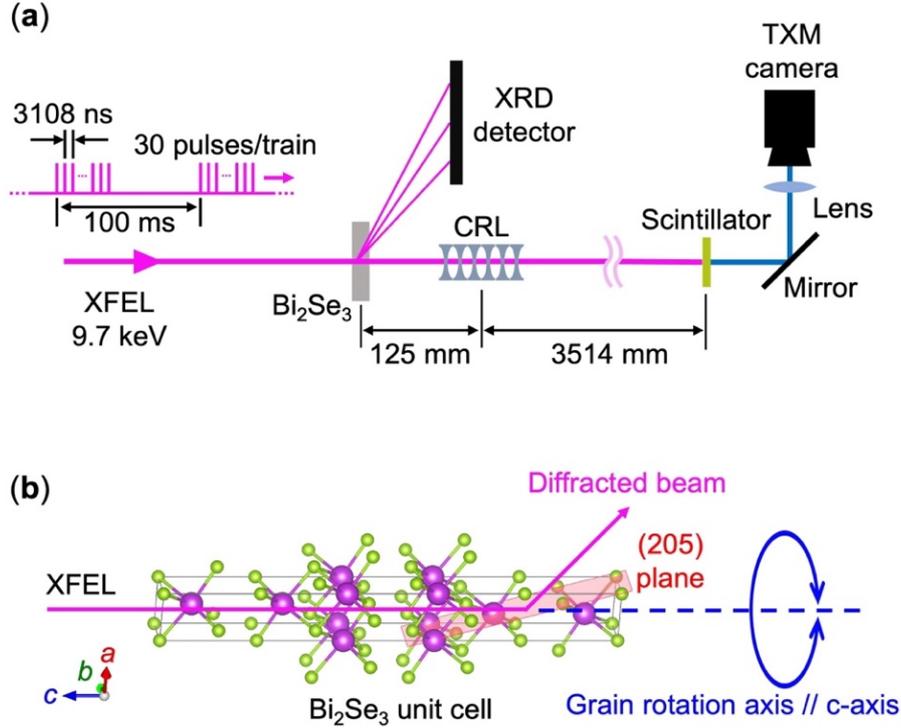

**Fig. 1.** (a) Schematic of the simultaneous TXM and XRD measurements used in this work at the HED instrument of European XFEL. Distances are not drawn to scale. (b) Schematic of the $Bi_2Se_3$ crystal structure [42]. The *c*-axis of $Bi_2Se_3$ (perpendicular to the basal plane) is nearly parallel to the incident X-ray beam direction and is collinear to the grain rotation axis. The purple and light green spheres represent Bi and Se atoms, respectively. The (205) diffraction plane of $Bi_2Se_3$ is shown as a red plane within the unit cell.

The X-ray pulses used in this study had a photon energy of $E = 9.7$ keV. The measured XFEL beam size on the $Bi_2Se_3$ sample was ~32-μm and ~52-μm size in horizontal and vertical directions, respectively, as defined by 4-jaw slits located ~10-m upstream from the sample. The collimated X-ray beam was comprised of a series of pulse trains arriving at a 10 Hz repetition rate (*i.e.*, 100 ms between pulse trains) (Fig. 1(a)). Each pulse train contained 30 individual ~30 fs X-ray pulses that were evenly spaced in time by a delay of 3108 ns between pulses.

For the XRD measurements, the X-ray beams diffracted by the sample were collected by using a 2-dimensional (2D) direct X-ray detector (Jungfrau, produced by the Paul Scherrer Institut [45,46]) placed 193 mm downstream of the sample. We placed a 300-µm thick aluminum foil in front of the XRD detector to attenuate the measured XRD signals and avoid detector saturation. For the TXM, we placed a stack of 50 2D beryllium CRLs ($R$ = 50-µm) along the transmitted beam to form a magnified image, using 300-µm and 400-µm pinholes at the entrance and exit windows of the CRL stack, respectively. The magnified image was then converted to visible light by a 35-µm thick LuAG:Ce scintillator, then further magnified with a 7.5x magnification infinity-corrected objective lens (Mitutoyo) before being recorded by a camera (Photonic Science sCMOS 4.2). Our spatial calibration using a 2000-mesh TEM grid indicates the total magnification achieved is 239x. While each TXM frame recorded the integration of all 30 pulses from a train, the shutter for our XRD measurements was timed to only integrate diffraction produced by the last 15 of those 30 pulses in each pulse-train. This strategy produced measured signals more interpretable in describing the heated volume where dislocation and GB motion are more active. Since each train was separated by 100 ms, both the XRD and TXM images were recorded at 10 frames per second (10 Hz frame rate).

A single crystal of $Bi_2Se_3$ with an initial thickness of 35 µm was used as the sample. The sample's surface normal was parallel to the basal plane of the crystal and was set nearly perpendicular to the direction of the incident X-ray beam. Our pre-characterization showed the initial structure of our $Bi_2Se_3$ crystal is $R$-$3m$, which is the known stable form of $Bi_2Se_3$ at ambient conditions [47]. The sample surface is parallel to the basal plane of the $Bi_2Se_3$. For XRD, we used the diffraction peak from the (205) planes of the $R$-$3m$ structure of $Bi_2Se_3$ to track the grain orientation, separation, and rotation. We selected (205) planes as the diffracting planes to track the grain rotation, as it is most efficient at probing small rotations of the crystal out of its initial (001) orientation while still satisfying the Bragg condition. This choice was based on the initial orientation of the $Bi_2Se_3$ crystal and the X-ray photon energy of 9.7 keV which provided high XFEL flux for the study. The scattering angle of the (205) peak position observed by XRD is at $2\theta$ = 43.8° and the theoretical angle between the prismatic and (205) planes of $Bi_2Se_3$ is $\omega$ = 19.9°. This means a slight modification of the sample placement angle: $2\theta/2 - \omega = 2.0°$ is required to satisfy the Bragg condition (See Fig. S2). The angle modification occurred by the X-ray induced heating as a strong and spotty XRD peak from the $Bi_2Se_3$ (205) planes suddenly appears during the X-ray irradiations, as discussed below and in Supplementary Movie S2. The initial angular offset of the sample was used to avoid saturating the XRD detector with the high-intensity diffraction from the initial pristine single-crystal peak. The quasi-2D layered atomic structure of $Bi_2Se_3$ crystal makes it easy to control the main grain rotation axis. This was predicted based on the van der Waals bonding between basal layers, and later confirmed by the XRD measurements, to be perpendicular to its basal plane. By making the grain rotation axis nearly collinear with the XFEL irradiation axis (Fig. 1(b)), we were able to track grain rotation as the change in the azimuthal angle of the diffracted peak. If the rotation was not along the X-ray irradiation axis, the XRD peak would have disappeared as the grain would rotate primarily in different directions that would no longer satisfy the Bragg condition.

In addition to these measurements, we performed temperature simulations based on the X-ray pulse energy applied to the crystal. Using the X-ray pulse energy measured upstream of the beamline, we estimate the pulse energy arriving at the front surface of the $Bi_2Se_3$ sample to be ~2.5 µJ/pulse. We used this energy to simulate the spatial distribution of temperature in the $Bi_2Se_3$ crystal induced by irradiation from one X-ray pulse train (30 pulses), as shown in Fig. 2(a). The details of the pulse energy estimation and simulations are included in Supplemental Material Sec III. Our simulation results suggest the temperature rise after irradiation of one X-ray pulse train would reach ~1600 K near the irradiated surface of the sample. As the melting temperature of $Bi_2Se_3$ is 978 K [48], we predict the X-ray irradiation caused the irradiated volume of the crystal to partially melt. The simulated temperature in $Bi_2Se_3$ also reaches the vaporization temperature which is estimated to be around 1600 K based on extrapolation of the pressure-dependent vaporization temperature data of [49], but the X-rays do not deliver enough energy to completely vaporize the crystal when considering the latent heat needed for the full vaporization in the simulation. Note that $Bi_2Se_3$ is known to have no high-temperature phase transitions up to its melting temperature at ambient pressure [48]. Our simulation also suggests that thermal diffusion between pulse trains carries the heat out of the hot spot in less than 100 ms (Fig. 2(b)). This indicates that the 3108 ns time separation over the full 30 X-ray pulses in each train is short enough to accumulate heat between pulses within a train at the crystal, while the 100 ms interval between each train is long enough to fully cool the system.

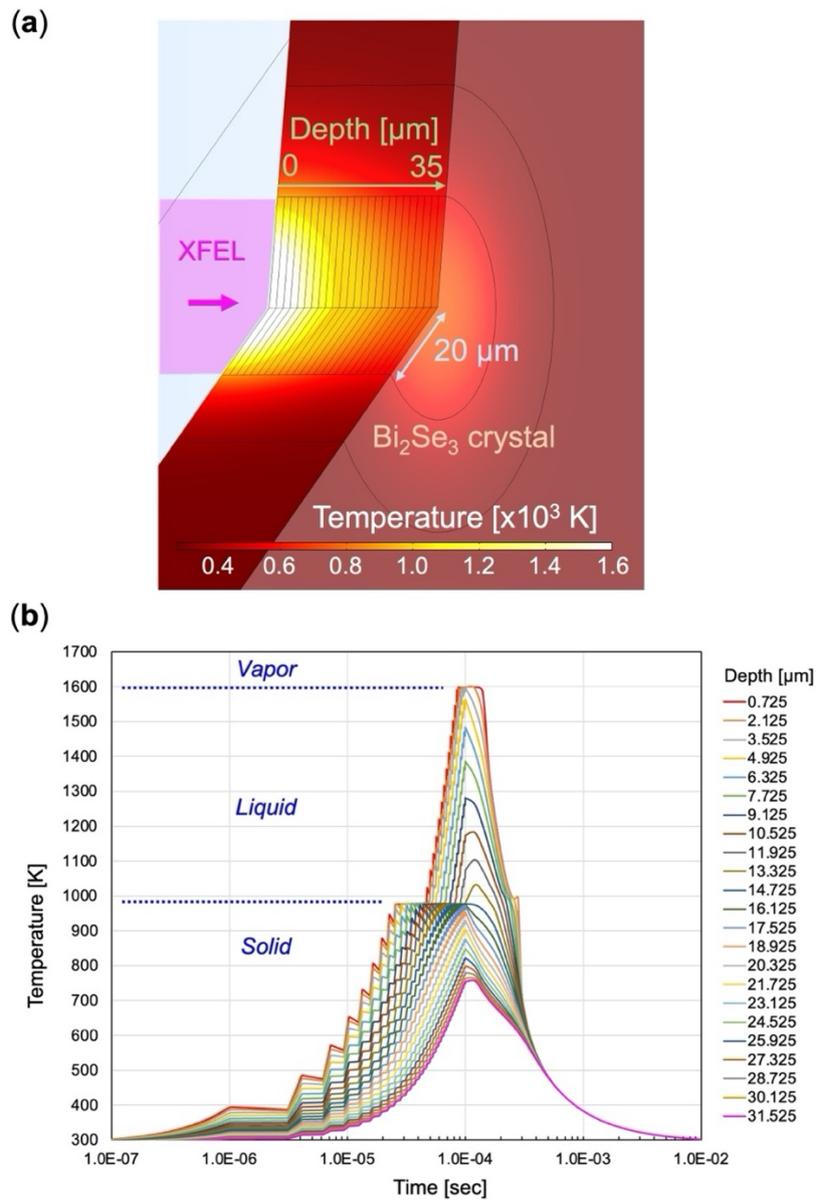

**Fig. 2.** Temperature simulations based on the estimated on-sample XFEL pulse energy. This simulation considers one pulse train that contains 30 XFEL pulses with a time spacing of 3108 ns between each pulse. (a) Simulated spatial temperature distribution at the end of the irradiation of one pulse train. The image has a radial cut-out slice to show the temperature profile inside the simulated volume. (b) Time dependence of the simulated temperatures at different depths from the XFEL irradiated side of the crystal surface.

Representative TXM images are displayed in Fig. 3 to show the time dependence of the X-ray-induced damage on the $Bi_2Se_3$ crystal. See Supplementary Movie S1 for the complete set of the recorded TXM images (0-20 s, 10 fps). At 14.9 s and after, a large hole can be observed in the center of the image, indicating that the X-ray-induced damage punctures fully through the crystal. Some of the TXM images also capture the formation of small circular features, as indicated by the red arrows in Fig. 3. The sizes of the features are up to several micrometers in diameter and once they appear, each feature remains for 1-3 image frames (0.1-0.3 s), at the same position without changing their sizes, before disappearing. We interpret these circular features to be bubbles resulting from the vaporized $Bi_2Se_3$ formed at the surface or trapped in the bulk of the crystal. Although our simulation (Fig. 2) indicates the maximum temperature in the crystal to be slightly lower than the vaporization temperature, the X-ray pulses we used had some spatial inhomogeneity and varied overall intensity between pulse trains (Supplemental material Sec IV) that could cause vaporization at local hotspots. We note that these bubble features tend to emerge in the brightest regions of our X-ray images, which is consistent with this interpretation.

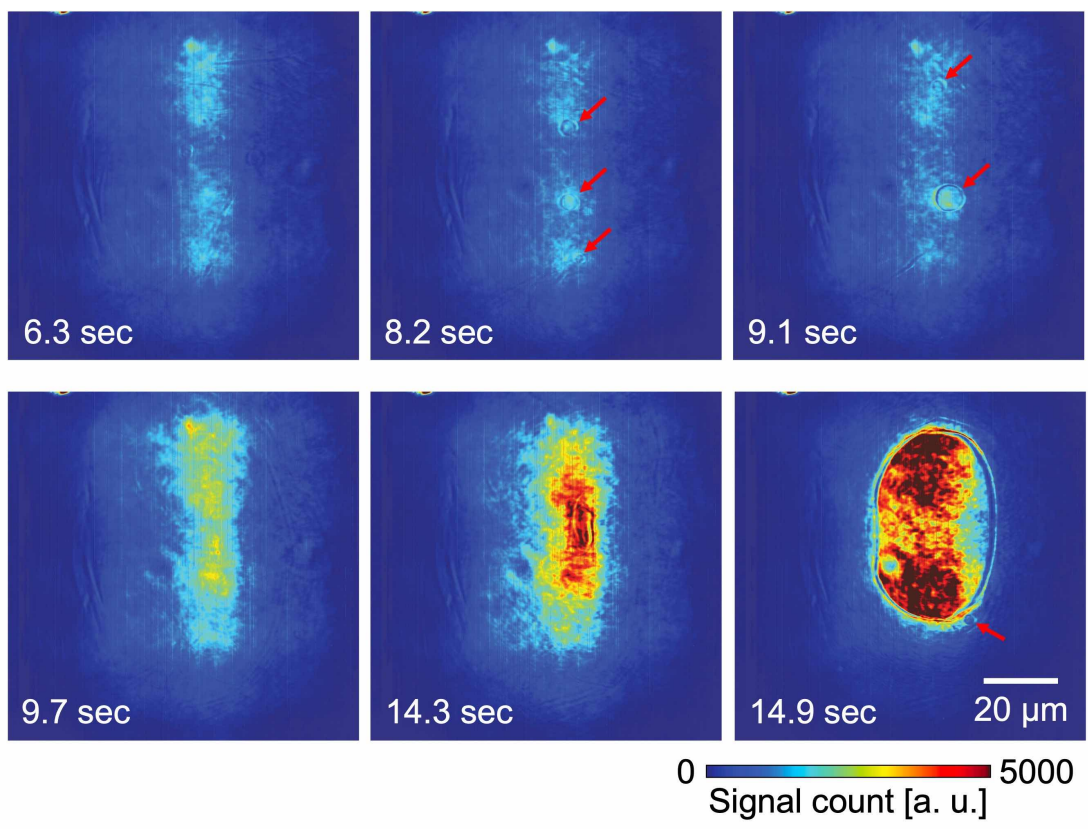

**Fig. 3.** TXM images showing time-dependence of the X-ray-induced damage on the Bi$_2$Se$_3$ crystal. Time 0 s denotes the arrival time of the first X-ray train irradiating the crystal. The observed signal counts increase as time progresses, as the X-ray irradiated volume of the crystal becomes thinner due to X-ray induced damage. A large puncture hole is observed in the images of 14.9 s and is retained through the end of the 20 s experiment. The red arrows indicate the appearance of the transient features we interpret to be bubbles. The scale bar on the 14.9 s image shows the sample dimensions measured at the imaging plane (*i.e.*, the scale at the sample, not at the detector).

Some of the recorded XRD images showing the time evolution of the GB formation and subsequent grain rotation in the $Bi_2Se_3$ crystal are shown in Fig. 4. See Supplementary Movie S2 also for the complete set of the recorded XRD images (0-20 s, 10 fps). The XRD images show how a spotty diffraction peak from the (205) plane observed at a scattering angle of $2\theta \cong 44°$ splits into three spots and then changes their azimuthal positions by the subsequent XFEL irradiation. The strong (205) diffraction peak observed at 8.1 s separates into three different peaks at 8.3 s, indicating one large grain splitting into smaller grains by forming subgrain boundaries (also known as low-angle grain boundaries). The misorientations of the two moving peaks at 8.3 s are $\Delta\varphi = 0.6°$ and -0.6°, respectively. The number of peaks was determined based on the existence of a local minimum by taking the second derivative of the azimuthal profile. The angular resolution of neighboring grains is 0.2°. The subsequent changes of the observed peak positions in azimuthal angle observed at 8.4 s and after show the grains rotating as they form high-angle GBs. We note that there should be many other grains formed during this experiment, though they would be invisible to our XRD measurement which can only detect grains aligned to the Bragg condition.

A closer look at the XRD peaks (Fig. 4) shows the broadening of the diffraction rings in $2\theta$ directions which is especially evident for the strongest diffraction peaks from the (205) planes. This should be the result of lattice expansion caused by temperature increases [50]. As we set each XRD image acquisition to accumulate the diffraction patterns from the last 15 XFEL pulses of the 30 pulses in each train, each XRD image records the sum of the diffraction signals from the same crystal but at 15 different states, each with slightly higher temperatures than the previous ones (Fig. 2(b)). Since the temperature increment causes the crystal to expand, the latest pulses should diffract to smaller $2\theta$ values than the earlier pulses, resulting in the observed broadening of the diffraction peaks in the $2\theta$ direction. While XRD peak width is often used to estimate the size of diffracting grains via the Scherrer equation, the above temperature effect makes the grain size estimation difficult. Also, the Scherrer equation is known to be valid for grains with average sizes up to around 0.1 μm. Our rough estimation of the size of the rotating grains is somewhere between sub-micrometer and tens of micrometers. The upper limit is set by the XFEL beam size (32 x 52 μm$^2$) and the lower limit is assumed based on the SEM analysis on a recovered sample that experienced a relatively lower X-ray dose (Fig. S6) showing formation of sub-micrometer grains on the irradiated surface.

While our temperature simulation suggests some fraction of the Bi$_2$Se$_3$ crystal to be melted under the X-ray induced heat, our XRD detected no sign of liquid scattering. This is thought to be because the volume fraction of the melted portion relative to the part remaining solid is always small as the melted and vaporized portions would not stay at the probed region unless they recrystallize rapidly. Additionally, the scattering intensity for liquids is much lower than for crystalline materials.

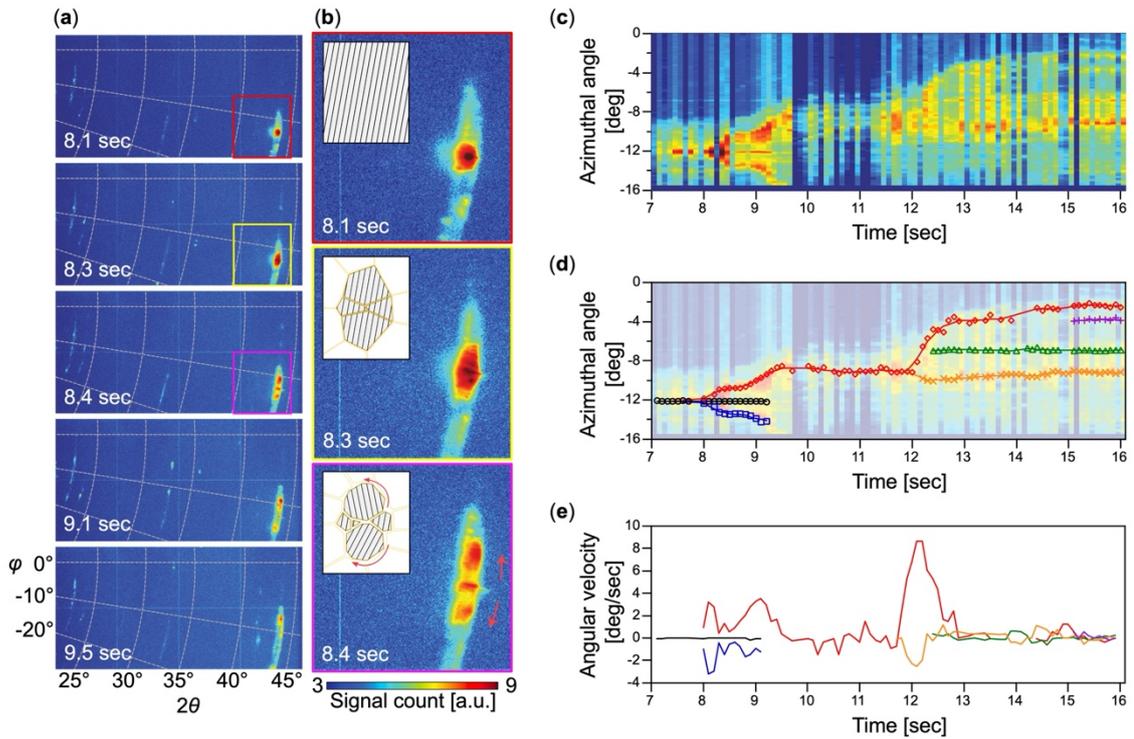

**Fig. 4.** GB formation and subsequent grain rotation of Bi$_2$Se$_3$ time-resolved by the XRD. The selected and noted times correspond to the measurements corresponding to the TXM images in Fig. 3. (a) XRD images taken at five representative times. The signal counts of the images are logarithmically scaled for clarity at the high dynamic range of the measurement. (b) Some of the images shown in (a), enlarged to show the detail of the XRD peak from (205) planes. Red arrows indicate the direction of the rotation along the azimuthal direction. Insets are schematics describing a possible explanation of the GB formation and subsequent grain rotation sequences observed by the XRD. (c) The azimuthal intensity distribution of the diffraction peak from (205) planes. (d) Traces of the XRD peak positions observed in (c). Plots are the measured

peak positions with different colors and shapes used for different traces. Solid lines are the moving average ($k = 5$) of the measured plots. (e) Angular velocities of each trace obtained by taking the derivative of the moving average traces shown in (d). The colors of the lines correspond to those in (d).

As the measured XRD time resolves the rotation of the grains with 10 fps, the speed of the grain rotation can be determined. Fig. 4(c) shows the azimuthal position of the (205) diffraction peak as a function of time and Fig. 4(d) shows the measured traces of the peak positions. By taking the derivative of the change of the azimuthal angles of the traces, we obtained the angular velocities of the $Bi_2Se_3$ grain rotations (Fig. 4(e)). The fastest grain rotation speed observed in this work reaches 8.6 deg/s. Huang *et al.* reported the grain rotation speed in nano-polycrystalline AgBr induced by irradiation of synchrotron X-rays reached 186 deg/s [7]. They also observed much slower grain rotation of AgBr with a speed comparable to our $Bi_2Se_3$ results (1-10 deg/s). The faster grain rotation speed observed by Huang *et al* should be due mainly to the effect of the smaller grain size of their crystal.

The grain fragmentation of nano-polycrystalline AgBr observed by Huang *et al.* [7] was induced by synchrotron X-ray beams focused to nanometers. They found that the grain fragmentation was induced by the photo-induced chemical decomposition of AgBr to Ag + Br. They described that the grain fragmentations drive the subsequent grain rotations. In our $Bi_2Se_3$ study, however, the XRD measurements did not show any sign of the decomposition of $Bi_2Se_3$ within its solid state, as we only measured the *R-3m* structure of the $Bi_2Se_3$ crystal (Fig. S1). Thus, we interpret the driver of the observed grain refinement of $Bi_2Se_3$ to be not the photo-induced chemical decomposition, but the X-ray induced thermal stress built up in the lattice exceeding the yielding stress of $Bi_2Se_3$. The subsequent grain rotation can be explained by the conventional understanding of dislocation-mediated grain rotation [51,52].

While the *in-situ* XRD did not show any sign of decomposition in the solid $Bi_2Se_3$, our scanning electron microscopy (SEM) and energy-dispersive X-ray spectrometry (EDS) analysis of the recovered $Bi_2Se_3$ sample observed decomposition of bismuth and selenium (Fig. 5). The observed selenium rich area around the hole indicates the non-diffracting liquid or vapor phase $Bi_2Se_3$ in the X-ray irradiated region (*i.e.*, within the puncture hole) decompose into bismuth and selenium during the irradiation, then spatter onto the unirradiated surface around the X-ray irradiated region. We note that any decomposed material solidified into the X-ray irradiated regions would also melt or vaporize again by following X-ray pulses. As we interpret the decomposition only occurs in the melted or vaporized volume, the decomposition would not directly affect the GBs.

To summarize, our results experimentally demonstrate that intense X-ray irradiation can rapidly induce GB formation and subsequent grain rotation in initially single crystalline $Bi_2Se_3$. We interpret the build-up of thermal stress induced by the X-ray irradiation as a possible mechanism for the observed change in GB characteristics. Although the high-intensity X-ray pulses used in this study caused melting and puncturing of the sample, such damage can be avoided by optimizing the irradiation conditions.

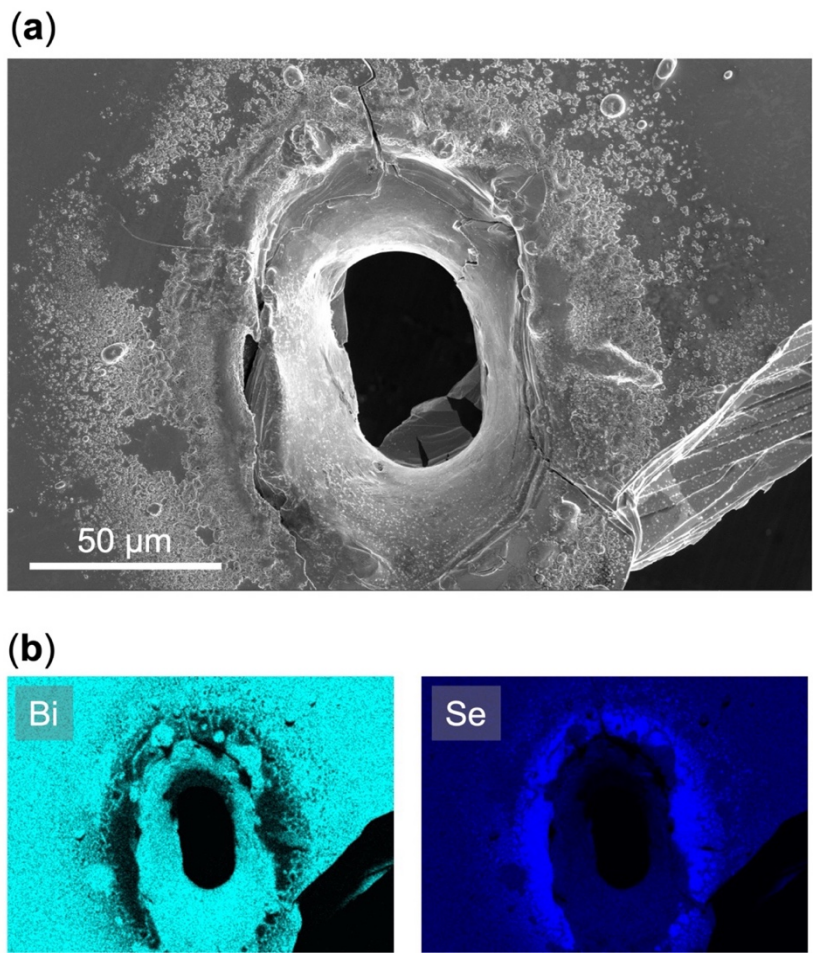

**Fig. 5.** Structure and elemental mapping of the recovered sample. (a) Scanning electron microscopy (SEM) image showing the puncture hole on the recovered $Bi_2Se_3$ sample after the 20 s experiment. A portion of the crystal (right-bottom part of the image) is missing as the sample was damaged during the SEM preparation process. (b) Energy-dispersive X-ray spectrometry (EDS) maps of bismuth (left) and selenium (right) showing the elemental segregation beyond the X-ray illuminated volume.

**CRediT authorship contribution statement**

**Kento Katagiri:** Writing – original draft, Formal analysis, Data curation. **Bernard Kozioziemski:** Writing – review & editing, Investigation, Methodology, Formal analysis, Data curation. **Eric Folsom:** Investigation, Methodology. **Sebastian Göde:** Investigation, Formal analysis. **Yifan Wang:** Writing – review & editing, Data curation. **Karen Appel:** Writing – review & editing, Investigation. **Darshan Chalise:** Writing – review & editing, Data curation. **Philip K. Cook:** Investigation. **Jon Eggert:** Investigation. **Marylesa Howard:** Investigation. **Sungwon Kim:** Investigation. **Zuzana Konôpková:** Investigation. **Mikako Makita:** Investigation. **Motoaki Nakatsutsumi:** Investigation. **Martin M. Nielsen:** Investigation. **Alexander Pelka:** Investigation. **Henning F. Poulsen:** Writing – review & editing, Investigation. **Thomas R. Preston:** Investigation. **Tharun Reddy:** Investigation. **Jan-Patrick Schwinkendorf:** Investigation. **Frank Seiboth:** Writing – review & editing, Investigation. **Hugh Simons:** Investigation. **Bihan Wang:** Investigation. **Wenge Yang:** Investigation. **Ulf Zastrau:** Investigation. **Hyunjung Kim:** Writing – review & editing, Investigation, Funding acquisition. **Leora E. Dresselhaus-Marais:** Writing – review & editing, Investigation, Funding acquisition, Methodology, Conceptualization, Methodology, Supervision, Project administration.

**Declaration of competing interest**

The authors declare that they have no known competing financial interests or personal relationships that could have appeared to influence the work reported in this paper.

**Acknowledgements**

We acknowledge European XFEL in Schenefeld, Germany, for the provision of X-ray free-electron laser beamtime at Scientific Instrument HED (High Energy Density science) and would like to thank the staff for their assistance. The authors are grateful to the HIBEF user consortium for the provision of staff that enabled this experiment. The original datasets can be found here and are available upon reasonable request: doi: 10.22003/XFEL.EU-DATA-002293-00. Characterizations of the initial sample and the recovered sample were performed at the Stanford Nano Shared Facilities (SNSF), supported by the National Science Foundation under award ECCS-2026822. Part of this work was performed under the auspices of the U.S. Department of Energy by Lawrence

Livermore National Laboratory (LLNL) under Contract DE-AC52-07NA27344. The initial experimental work in this study was supported by the Lawrence Fellowship at LLNL. This work was done in part by Mission Support and Test Services, LLC, under Contract No. DE-NA0003624 with the U.S. Department of Energy, the National Nuclear Security Administration. DOE/NV/03624--2000. S.K. and H.K. acknowledge the support of the National Research Foundation of Korea grant NRF-2021R1A3B1077076. The authors declare no conflict of interest.

**Supplementary materials**

Supplementary file and three supplementary movies are available.

Supplemental Material

# X-ray induced grain boundary formation and grain rotation in $Bi_2Se_3$

Kento Katagiri[1,2,3*], Bernard Kozioziemski[4], Eric Folsom[4], Sebastian Göde[5], Yifan Wang[1,2,3], Karen Appel[5], Darshan Chalise[1,2,3], Philip K. Cook[6], Jon Eggert[4], Marylesa Howard[7], Sungwon Kim[8], Zuzana Konôpková[5], Mikako Makita[5], Motoaki Nakatsutsumi[5], Martin M. Nielsen[9], Alexander Pelka[10], Henning F. Poulsen[9], Thomas R. Preston[5], Tharun Reddy[1,2,3], Jan-Patrick Schwinkendorf[5], Frank Seiboth[11], Hugh Simons[9], Bihan Wang[12], Wenge Yang[12], Ulf Zastrau[5], Hyunjung Kim[8], and Leora E. Dresselhaus-Marais[1,2,3,4*]

[1]Department of Materials Science and Engineering, Stanford University, Stanford, CA 94305, USA

[2]SLAC National Accelerator Laboratory, Stanford, CA 94025, USA

[3]PULSE Institute, Stanford University, Stanford, CA 94305, USA

[4]Lawrence Livermore National Laboratory, Livermore, CA 94550, USA

[5]European XFEL GmbH, Schenefeld 22869, Germany

[6]Institute for Physics and Materials Science, University of Natural Resources and Life Sciences (BOKU), Vienna 1190, Austria

[7]Nevada National Security Site, North Las Vegas, NV 89030, USA

[8]Department of Physics, Sogang University, Seoul 04107, Korea

[9] Department of Physics, Technical University of Denmark, Kgs. Lyngby 2800, Denmark

[10]Helmholtz-Zentrum Dresden-Rossendorf (HIBEF), Dresden D-01328, Germany

[11]Center for X-ray and Nano Science CXNS, Deutsches Elektronen-Synchrotron DESY, Hamburg 22607, Germany

[12]Center for High-Pressure Science and Technology Advanced Research, Shanghai 201203, China

*Corresponding authors. Email addresses:

kentok@stanford.edu (K. Katagiri),

leoradm@stanford.edu (L. Dresselhaus-Marais)

# I. Structure of the Bi$_2$Se$_3$ sample under X-ray irradiation

The observed structure of the Bi$_2$Se$_3$ sample under X-ray irradiation is *R-3m* (Figure S1), which is the stable phase of Bi$_2$Se$_3$ at ambient conditions. Our *in-situ* XRD measurements did not show any signs of phase transformations or decomposition into Bi and Se. The XRD profiles collected at 9.5 s (Figure S1) suggest the peak from (205) is much stronger than other peaks, indicating the grain rotation is mostly along the *c*-axis (normal to the basal plane).

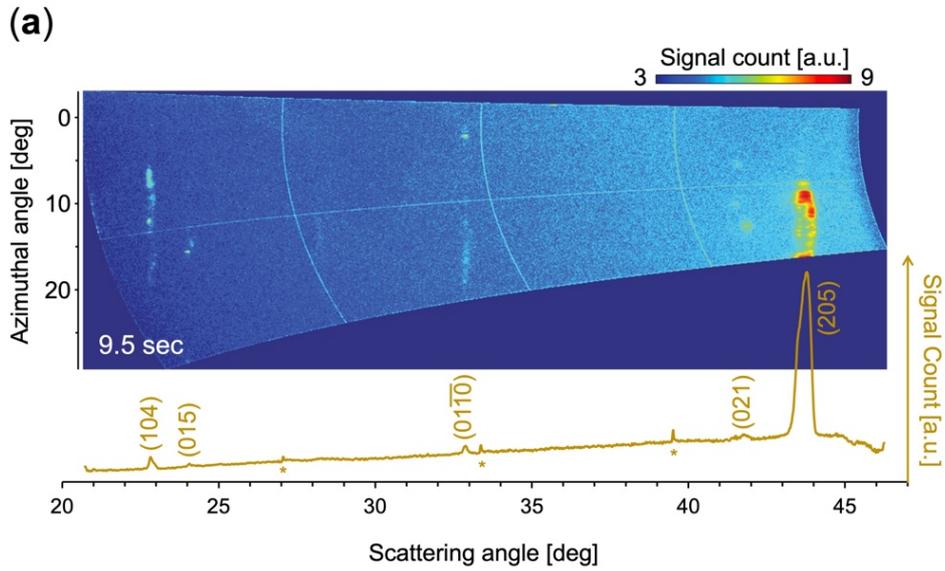

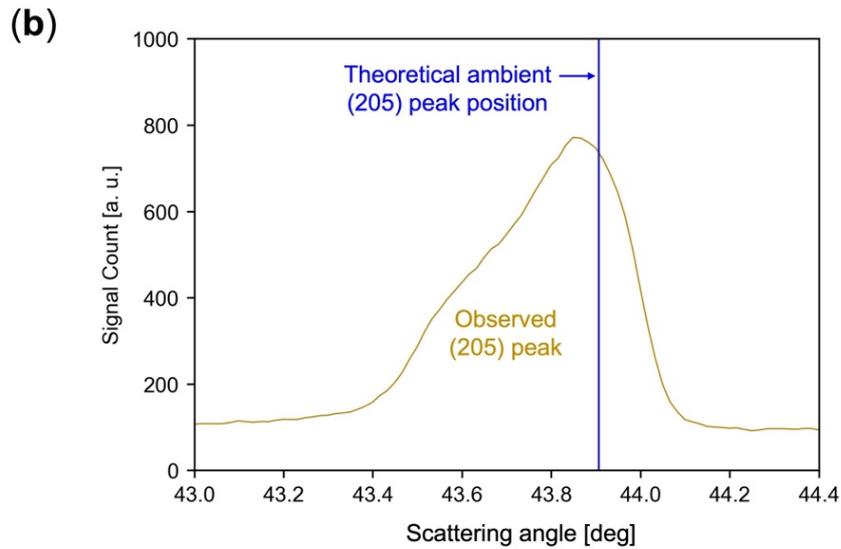

**Fig. S1.** (**a**) A polar-transformed version of the XRD image acquired at 9.5 s (top) with a plot of the corresponding line profile after azimuthal integration (bottom). The XRD image has a color bar with an intensity that is logarithmically scaled for image clarity, while the intensity of the corresponding line profile is scaled linearly to convey its interpretation. All the crystal planes denoted on the line profile are for the *R-3m* structure of the $Bi_2Se_3$. The small and sharp peaks in the trace that are denoted with * correspond to artifacts from junctions between different panels of the detector. (**b**) Detail of the XRD profile near the (205) peak. The blue line indicates the diffraction peak position predicted for the (205) plane of $Bi_2Se_3$ at ambient conditions, as estimated by assuming lattice constants of *a* = 4.143 Å, *b* = 4.143 Å, and *c* = 28.636 Å as measured in ref [53]. The experiment observed the (205) diffraction peak at lower scattering angle (2θ) due to thermal expansion from the X-ray irradiation.

## II. Satisfying the Bragg condition for the (205) planes

As described in the main text, the basal plane of the $Bi_2Se_3$ sample was set to be perpendicular (normal) to the axis of the incoming (incident) XFEL beam (Fig. S2(a)). The initial orientation of the single crystal does not meet the Bragg condition for the (205) plane. From the geometric calculations shown in Fig. S2, for the new grains of the crystal to meet the Bragg condition for the (205) planes, they must rotate by ~2.0°, as depicted in Fig. S2(b). The initial slight offset of the angle from the ideal Bragg condition was used to avoid saturating the XRD detector with the highly intense diffraction beams produced by pristine single-crystals, which are known to damage sensitive detectors like the one used in this work. The sudden appearance of the (205) peak on the XRD detector at ~7 sec (see Movie S2) thus indicates the 2.0° rotation had occurred as a result of the XFEL irradiation. As described in the main text, we interpret this to be because the X-ray induced temperature rise caused anisotropic thermal expansion and slight grain rotations.

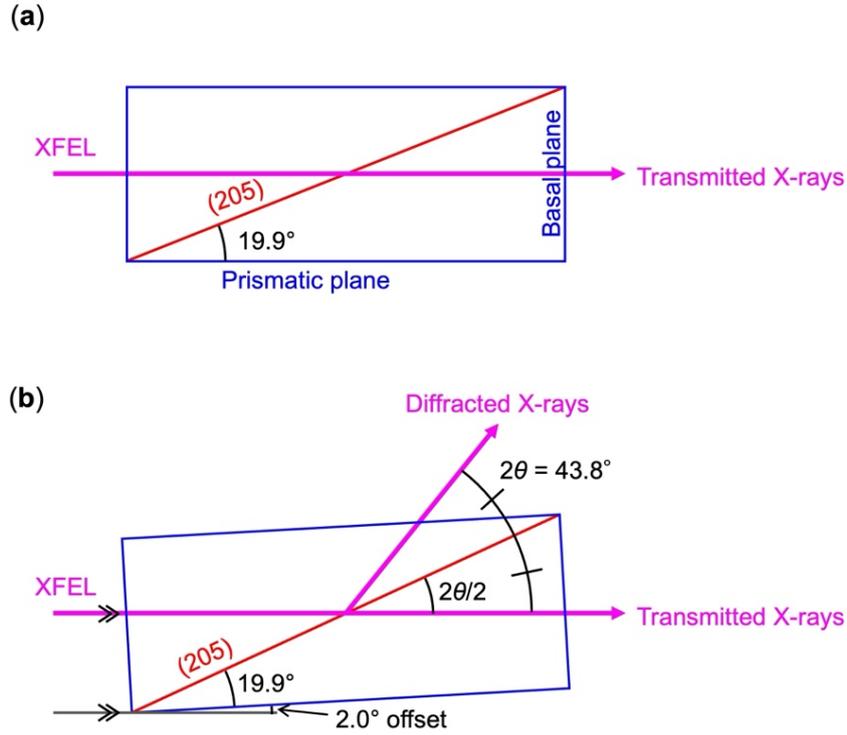

**Fig. S2.** Drawings showing the diffraction plane angle (red) relative to the XFEL irradiation direction (magenta). Blue rectangle is shown to indicate the basal and prismatic planes of the $Bi_2Se_3$ sample. Figures not drawn to scale. (**a**) The angular relationship when the X-ray incidence angle is set normal to the $Bi_2Se_3$ basal plane, for which the Bragg condition of the (205) planes is not satisfied and no XRD results. (**b**) The angular relationship between the (205) plane's normal vector and the incident X-ray beam when the Bragg condition is satisfied for the (205) planes due to the 2.0° offset.

## III. Temperature simulation based on the estimated XFEL pulse energy on the sample

In our experiments, the pulse energy of the XFEL was monitored by using an X-ray gas monitor (XGM) [54] located upstream of the sample. For the calibration runs, the average pulse energy XGM measured with the attenuator set to 29% transmission was as in the experiment was ~120 µJ/pulse. For the run we presented in this work, we increased the transmission through the attenuator to 75%. By assuming the pulse energy before the attenuator to be the same between the calibration and measurement runs, we estimate 120×0.75/0.29 = 310 µJ/pulse without attenuation. As 97% of the beam produced by the accelerator was blocked by the 4-jaw slits located between the XGM and the sample (to ensure beam position stability), the pulse energy at the sample position would correspond to 8.7 µJ/pulse. Considering the additional attenuation from air and diamond windows in the beamline infrastructure, we estimate 2.5 µJ/pulse at the front surface of the $Bi_2Se_3$ sample.

The X-ray induced temperature rise in the 35µm thick $Bi_2Se_3$ was then simulated using COMSOL Multiphysics® software [55], based on the estimated X-ray pulse energy onto the sample. The XFEL irradiation area was set as a 40 µm diameter circle in the simulation (performed as an axisymmetric simulation to enhance computational efficiency), which gives a beam size comparable to the measured beam size on the sample (32 x 52 µm$^2$). The X-ray energy absorbed per pulse was estimated using LBNL's CXRO database [56], for positions along the depth of the sample that were split into a grid size of 2.5 µm through the depth of the sample. We use a thermal conductivity value of 2 W/mK, which are the average of through-plane and in-plane conductivities at 300 K [57,58]. The specific heat capacity and density of $Bi_2Se_3$ used are room temperature values of 190 J/(kg•K) and 6.8 g/cm$^3$, respectively [48,59]. The enthalpy of melting and vaporization are estimated to be 131x10$^3$ J/kg [59] and 320 kJ/kg [48,49,60], respectively. The simulated temperature profiles are shown in Figure 2 of the main text.

Our temperature simulations suggest there is a non-negligible amount of thermal diffusion that heats the crystal outside of the X-ray irradiated volume (Fig. 2a). Our XRD captured diffracted signal from the $Bi_2Se_3$ sample even after the X-ray beam punctured most of the irradiated sample. This implies that some dynamics in the grain boundaries and grain rotation continue to occur outside the region illuminated by the brightest region of the beam. Our TXM also suggests a non-zero X-ray intensity outside the limits of the main beam, which would contribute to the XRD signals even though those X-rays may not be strong enough to drive the grain rotation dynamics. While the XRD signal contributions from inside and outside the main beam cannot be separated in our experimental geometry, as they are both part of the gauge volume that defines the XRD integration region, we interpret the dynamics of large grains observed at early stages of irradiation would only occur at the volume irradiated by the main beam (*i.e.*, inside the volume punctured at 14.9 s).

**IV. Characterization of the X-ray beam**

To characterize the spatial pattern of the X-ray beam, we recorded TXM images without a sample. A typical image is shown in Fig. S3 and a movie containing 100 images recorded at 10 fps is shown in Movie S3. The observed spatial intensity distributions of each frame in the movie suggest that the position of the beam was stable, while the beam profile has some spatial inhomogeneity. The movie also suggests there are some intensity fluctuation between subsequent pulse trains, which is quantitatively evaluated in Fig. S4. We again note that each TXM frame accumulated X-ray signals of 1 pulse train which contains 30 X-ray pulses.

As discussed in the main paper, our temperature simulations based on the average X-ray intensity predict the peak temperature in the $Bi_2Se_3$ sample to be slightly lower than the vaporization temperature (Fig. 2). However, our TXM observations show transient formations of bubbles within the X-ray irradiated volume of the $Bi_2Se_3$ crystal, suggesting partial vaporization of the crystal. This can be explained by the local hot-spots of the X-ray beam pattern and the train-by-train fluctuation of the X-ray intensities, making local X-ray induced temperature rise higher than what our simulation estimated based on the average X-ray intensity.

While the images in Movie S3 illustrate the heterogeneity in the spatial mode of the XFEL beam used in this work and the train-by-train intensity fluctuations, we note that they differ from the actual shape used in the run from the main text. The slit were open for the calibration run (Movie S3) while it was used to limit the beam size for the main experiment (Movie S1) – generating different beam sizes and shapes due to diffractive effects of the optical system. We include Movie S3 to demonstrate the extent of the fluctuations that give rise to hotspots in our material, as those trends are not sensitive to the slit sizes.

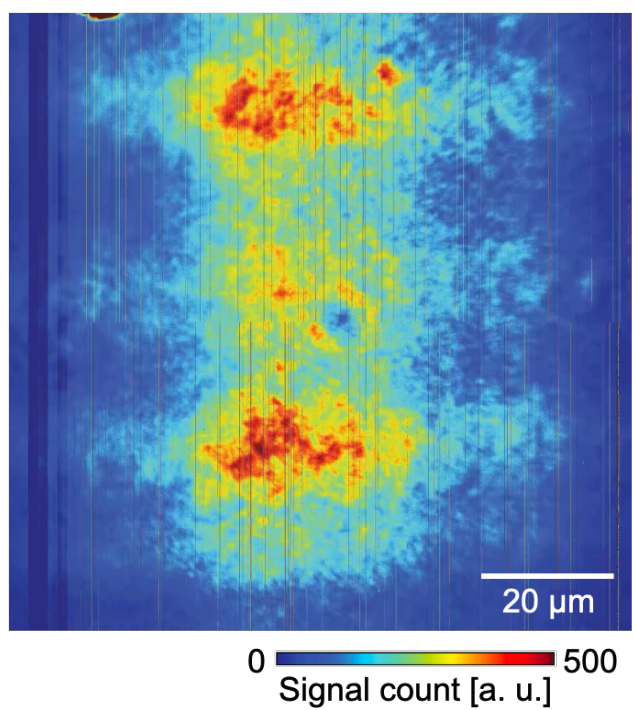

**Fig. S3.** A typical TXM image recorded with no sample mounted. The 4-jew slits were open but the CRLs were in to magnify the X-ray beam. The signal counts are lower than those observed in Fig. 3 because more attenuators were used for this calibration run.

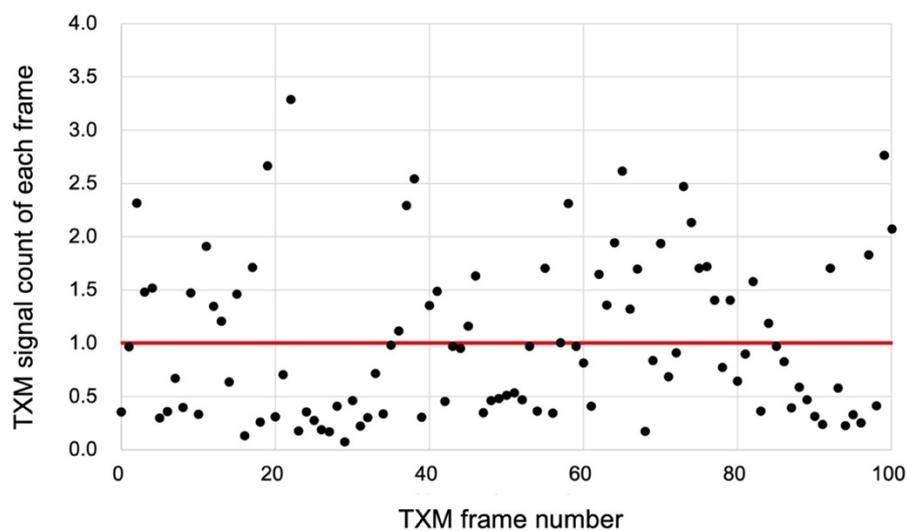

**Fig. S4.** Fluctuation of the measured X-ray pulse trains, as calibrated by integration of the TXM pixel intensities from Movie S3 that was collected without a sample. Each black point shows a signal of a TXM image divided by the average of 100 data points (red line).

**V. "Tilted" vs "Twisted" grain rotation**

In this section, we discuss two possible grain rotation mechanisms: "Tilted" and "Twisted" grain rotations depicted in Fig. S5.

Immediately after the (205) XRD peak split into the three (or more) domains (Figure 4), two of the three diffraction peaks rotated in opposite directions at a similar speed (see red and blue profiles at ~8-9 s in Figure 4e). This indicates that the temperature conditions that drive the grain rotations are similar for these grains and suggests that they are located at the same depth within the sample. If the crystal is experiencing "twisted" grain rotations, the grain closer to the surface is at a higher temperature and thus expected to be rotating faster than the other grain. Thus, we believe the observed grain rotations of $Bi_2Se_3$ is more likely "tilted" grain rotations rather than a "twisted" grain rotations [61], if not a combination of both tilt and twist.

It is important to note that the grain boundary formation between the van der Waals bonded layers should be lower energy and thus easier to cleave than the bonded structures along the out-of-plane axis. We note, however, that the full grain formation mechanism would still require grain boundary formation along the plane perpendicular to the layers because the initial crystal is larger than the XFEL beam size. We thus interpret that the low-energy plane may initiate this process, but that the rotation of the crystal that the energy barrier of grain boundary formation along the higher-energy axis perpendicular to the layers would follow at a slower rate. We note that dislocation activities may facilitate this mechanism, though that analysis is beyond the scope of our measurements or models in this work.

To better understand of the structural evolution of the $Bi_2Se_3$ under the X-ray irradiation, we performed SEM observation on a different position of the $Bi_2Se_3$ sample where different X-ray irradiation conditions were achieved (Fig. S6). For this run, we used an additional X-ray attenuator, resulting in an X-ray intensity that was three times lower than the 20-second-long run presented in the main paper. Except the images shown in Figure S6 and the discussion in this paragraph, the data shown and discussed in the main paper and this supplemental information are all from the 20-second-long run which punctured the $Bi_2Se_3$ sample. Because of the weaker X-ray intensity and resulting slower damage speed for the additional run, we were able to stop the X-ray beam before the sample got punctured. The SEM images presented in Fig. S6 show sub-micrometer sized grains on the X-ray irradiated surface of the crystal. The SEM image suggests the grains are randomly oriented, thought to be the results of the grain rotations.

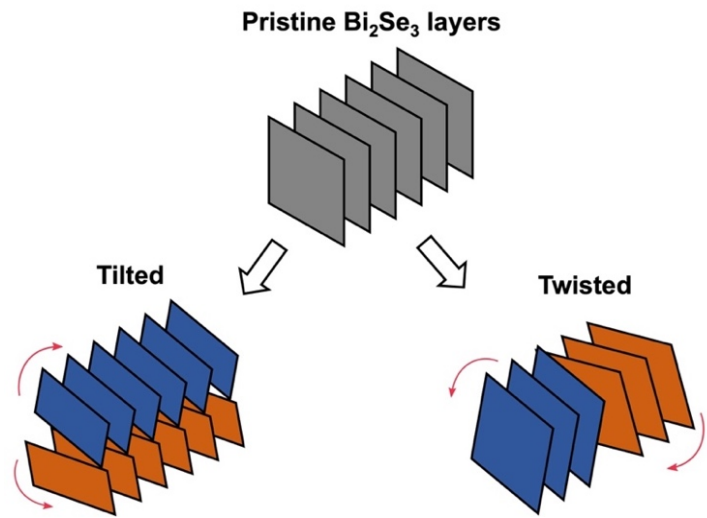

**Fig. S5.** Two possible explanations for the observed grain boundary formation and grain rotations. Each plane shows a layer of $Bi_2Se_3$. Rotating plans are colored blue and orange for better visibility. Rotation directions are indicated by the red arrows. The insets in Figure 4b also illustrate the tilted grain rotations we interpret for this system.

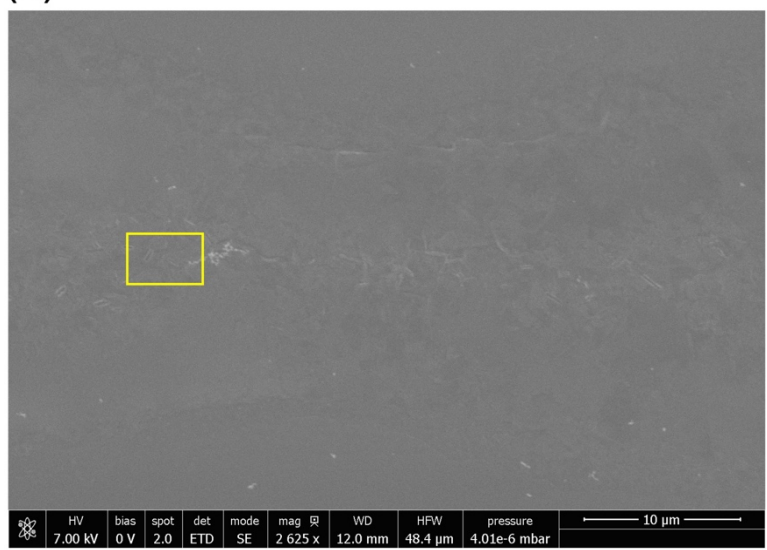

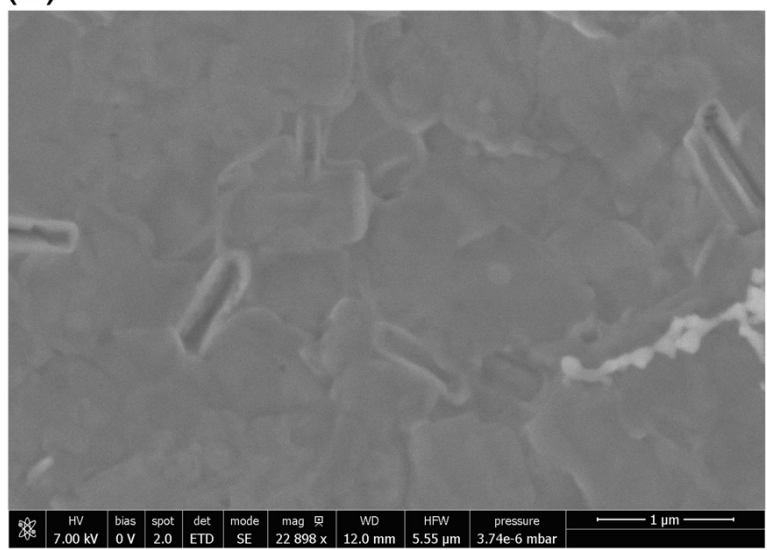

**Fig. S6.** (**a**) SEM image of $Bi_2Se_3$ sample. The X-ray intensity for this run was three times weaker than that for the 20 s run presented in the main paper (see text). (**b**) Enlarged image of (a), showing sub-micron sized grains formed by the X-ray irradiation. The yellow rectangle in (a) indicates the size and location where the image shown in (b) is acquired.